\documentclass[prd,twocolumn,showpacs,nofootinbib]{revtex4}
\usepackage{graphicx}
\usepackage{amsmath}
\usepackage{wasysym}
\usepackage{bm}
\begin{document}

\newcommand{\mwimp}{$m_\chi$}
\newcommand{\sigmapsi}{$\sigma_p^{SI}$}
\newcommand{\sigmapsd}{$\sigma_p^{SD}$}
\newcommand{\sigmansd}{$\sigma_n^{SD}$}
\newcommand{\kms}{km~s$^{-1}$}
\newcommand{\vsun}{$v_\odot$}
\newcommand{\ve}{$v_\oplus$}
\newcommand{\vescsun}{$v^{\odot}_{esc}$}
\newcommand{\vesce}{$v^{\oplus}_{esc}$}
\newcommand{\fjup}{$f_\jupiter$}
\newcommand{\fhalf}{$f_>$}

\title{Dark matter in the solar system II: WIMP annihilation rates in the Sun.}
\author{Annika H. G. Peter}
\email{apeter@astro.caltech.edu}
\affiliation{Department of Physics, Princeton University, Princeton, NJ 08544, USA}
\affiliation{California Institute of Technology, Mail Code 105-24, Pasadena, CA 91125, USA}
\date{\today}

\begin{abstract}
We calculate the annihilation rate of weakly interacting massive particles (WIMPs) in the Sun as a function of their mass and elastic scattering cross section.  One byproduct of the annihilation, muon neutrinos, may be observed by the next generation of neutrino telescopes.  Previous estimates of the annihilation rate assumed that any WIMPs from the Galactic dark halo that are captured in the Sun by elastic scattering off solar nuclei quickly reach thermal equilibrium in the Sun.  We show that the optical depth of the Sun to WIMPs and the gravitational forces from planets both serve to decrease the annihilation rate below these estimates.  While we find that the sensitivity of upcoming km$^3$-scale neutrino telescopes to $\sim 100\hbox{ GeV}$ WIMPs is virtually unchanged from previous estimates, the sensitivity of these experiments to $\sim 10\hbox{ TeV}$ WIMPs may be an order of magnitude less than the standard calculations would suggest.  The new estimates of the annihilation rates should guide future experiment design and improve the mapping from neutrino event rates to WIMP parameter space.

\end{abstract}


\maketitle

\section{Introduction}\label{sec:intro}
There is overwhelming evidence that non-baryonic dark matter must exist in large quantities in the universe, yet its composition remains unknown.  An interesting possibility, especially in light of anticipated data from the Large Hadron Collider (LHC) and the \emph{Fermi Gamma-Ray Space Telescope}, is that dark matter consists of at least one species of elementary particle that is part of an extension to the Standard Model (SM) of particle physics.  Such extensions bridge the theoretical gap between the electroweak symmetry breaking scale ($\sim 10^2$ GeV) and the Planck scale ($\sim 10^{19}$ GeV), and many naturally produce WIMPs in roughly the quantities and with behavior consistent with observations \cite{spergel2007} (e.g., the neutralino in supersymmetry \cite{jungman1996}, the Kaluza-Klein photon in universal extra dimension (UED) models \cite{cheng2002,servant2002,hooper2007}, or the heavy photon in Little Higgs models \cite{hubisz2005}).

There are three main approaches to the detection and characterization of WIMPs.  First, WIMPs may be created at the LHC or in future collider experiments \cite{baltz2006}.  Secondly, astrophysical WIMPs may be directly detected by measuring the energy deposited in a target during an interaction with a nucleus.  While dark matter has not yet been conclusively detected in such experiments (but see \cite{bernabei2000a,bernabei2008}), limits on WIMP-baryon cross sections are becoming ever more stringent.  Current experiments have target masses of $\sim 10$ kg, and are approaching sensitivities to elastic spin-independent WIMP-proton cross sections of \sigmapsi$\sim 10^{-44}$ cm$^2$ \cite{angle2008,cdms2008,hooper2007c}.  The experiments are starting to cut through swaths of minimal supersymmetric standard model (MSSM) and Little Higgs parameter space, but have yet to reach the cross sections possible with the simplest UED models \cite{hooper2007c}.  In the near future, kiloton-sized experiments should probe spin-independent cross sections down to $\sigma_p^{SI} \sim 10^{-46}$ cm$^2$ \cite{hime2006,gaitskell2007,akerib2008}.  WIMPs can also have spin-dependent interactions with nuclei, but experimental limits on those are much weaker than on spin-independent cross sections \cite{angle2008b,aubin2007,lee2007,behnke2008}.  

Thirdly, it may be possible to detect particles created in annihilations of WIMP pairs.  Many experiments (most obviously \emph{Fermi}, but see also \cite{tsuchiya2004,aharonian2006,bergstrom2006,antares2007b,superk2008,pamela2008,chang2008}) are searching for annihilation products from the Galactic center, the Milky Way halo, or Local Group satellite galaxies---the latter sites are promising because they are predicted to have high dark matter densities, and the annihilation rate goes as the square of the density.

The Sun may also be a source of high energy neutrinos, arising from WIMP annihilations at its core, since its potential well is deep enough that WIMPs may be captured (i.e., scattered onto orbits with speeds less than the local escape speed from the Sun) from the Galactic dark halo by elastic scattering off solar nuclei.  This signal may be observed by the next generation of large neutrino telescopes \cite{amram1999,icecube2001,dewolf2008}.\footnote{The Earth may also accumulate WIMPs \cite{gould1987}.  However, we have shown \cite{peter2009} that the neutrino event rates for WIMP masses and cross sections not yet excluded by direct detection experiments fall far below the thresholds for current or planned experiments.}  For the locus of models that are consistent with direct detection constraints and may produce an observable WIMP annihilation flux of neutrinos, the capture rate of WIMPs in the Sun is dominated by spin-dependent interactions.  In particular, since most of the mass of the Sun is in the form of hydrogen, and hydrogen is the only species in the Sun with a significant spin-dependent capture rate, neutrino telescopes should have very good sensitivity to \sigmapsd, the spin-dependent WIMP-proton cross section, currently the most poorly constrained of all WIMP-baryon elastic scattering cross sections.

The power of neutrino telescopes to detect solar WIMP annihilation has been demonstrated by the Super-Kamiokande experiment.  The upper bound on the neutrino flux from this experiment was used to determine a conservative upper limit to \sigmapsd$\,$ as a function of WIMP mass (using a method described in \cite{kamionkowski1995}) that is an order of magnitude better than the best limit set by direct detection, assuming that the WIMP is the supersymmetric neutralino \cite{desai2004}.  The next generation of neutrino telescopes will have much larger detector areas ($0.1$ km$^2$ for Antares \cite{amram1999}, 1 km$^2$ for IceCube \cite{icecube2001} and KM3NET \cite{dewolf2008}, versus $\sim 10^{-3}$ km$^2$ for Super-Kamiokande), albeit with somewhat higher energy thresholds, and should be sensitive to much lower elastic scattering cross sections than Super-Kamiokande (by two orders of magnitude for IceCube \cite{delosheros2008}).

To estimate the event rate in neutrino telescopes from WIMP annihilation in the Sun for a particular model, one needs three basic ingredients.  (i) An estimate of the  number density $n(\mathbf{r})$ of WIMPs in the Sun, and hence, the annihilation rate $\Gamma_a \propto n^2$.  (ii) An understanding of how the decay products, especially neutrinos and particles that decay to neutrinos after having traveled some distance in the Sun, interact and propagate through the Sun, interplanetary space, and the Earth.  (iii) A model of the charged-current interactions that produce muons in and near the detector volume and the sensitivity of the telescope to those muons.  Thus, the event rate in a neutrino telescope can be described schematically by
\begin{eqnarray}
	\Phi \sim \Gamma_a \times \binom{\hbox{neutrino}}{\hbox{physics}} \times \binom{\hbox{detector}}{\hbox{response}}.
\end{eqnarray}

Ingredient (iii) is studied as part of the neutrino telescope design process.  There has recently been significant progress in understanding ingredient (ii) \cite{cirelli2005,barger2007b,blennow2008,lehnert2008}.  One consequence of the work on this subject is that the muon neutrino flux at the Earth is quite uncertain for fixed WIMP mass and interaction cross sections.  Neutrino oscillations may enhance or suppress the relevant $\nu_\mu$ flux in neutrino telescopes depending on the annihilation channel.  For a given WIMP mass and annihilation rate in the Sun, the event rate at the Earth can vary by a factor of $\sim 10$ depending on the annihilation branching ratios.  

Traditionally, ingredient (i) has been treated as the simplest of these three sub-tasks in estimating neutrino event rates.  In the standard picture, hereafter called the ``instantaneous thermalization model'', WIMPs are captured in the Sun via elastic scattering off solar nuclei, and then quickly thermalize into a dense core at the very center of the Sun.  The thermalization is crucial to generating a large annihilation rate since the annihilation rate varies as the square of the density of the core.  This model of rapid thermalization of all captured heavy WIMPs is oversimplified for two reasons.  First, the thermalization time depends on both the  WIMP-nucleon cross section and mass, since the cross section governs the characteristic time between scatters and the mass governs the number of scatters required for WIMPs to reach equilibrium with solar nuclei.  If either the elastic scattering cross section is small or the WIMP mass is large, thermalization timescales can be long.  Secondly, the Sun is surrounded by a complement of planets.  The gravitational torques from these planets can change---or even reduce to zero---the frequency with which WIMPs encounter solar nuclei by altering the orbital paths of the WIMPs through the solar system.

The goal of this paper is to investigate the validity of the instantaneous thermalization model and to provide more accurate estimates of the WIMP annihilation rate in the Sun.  Our principal tool will be a large set of numerical simulations of the evolution of WIMPs captured in the Sun, including the gravitational influence of the planets and rescattering by solar nucleons \cite[][hereafter Paper I]{peter2009}.  We will demonstrate that in some cases the instantaneous thermalization model overestimates the annihilation rate by a factor of ten or more.

In Section \ref{sec:std}, we describe the standard instantaneous thermalization calculation of the annihilation rate of WIMPs in the Sun.  In Section \ref{sec:sim}, we briefly describe our simulations, which are discussed in more detail in Paper I.  We show how the results of the simulations modify the standard picture of WIMP annihilation in Section \ref{sec:insight}, and discuss our results in Section \ref{sec:conclusion}.

\section{The Instantaneous Thermalization Model}\label{sec:std}
The annihilation rate of WIMPs in the Sun is given by
\begin{eqnarray}\label{eq:gamma_a}
	\Gamma_a (t) = \langle \sigma v \rangle_a \int \text{d}^3\mathbf{r}\, n^2(\mathbf{r},t),
\end{eqnarray}
where $n(\mathbf{r},t)$ is the time-dependent number density of WIMPs in the Sun, and $\langle \sigma v\rangle_a$ is the velocity-averaged annihilation cross section.  For the remainder of this work, we take $n(\mathbf{r})$ to be the number density of WIMPs captured in the Sun, since in all cases where WIMP annihilation is significant, the annihilation rate due to captured WIMPs is many orders of magnitude greater than the rate due to halo dark matter streaming through the solar system.

If the captured WIMP number density can be separated as
\begin{eqnarray}\label{eq:num_sep}
	n(\mathbf{r},t) = N(t)\tilde{n}(\mathbf{r}), 
\end{eqnarray}
an assumption that we show below to be valid for the instantaneous thermalization model, and $\tilde{n}(\mathbf{r})$ is known and normalized such that $\int \tilde{n}(\mathbf{r})\text{d}^3\mathbf{r} = 1$, then the number $N$ of WIMPs in the Sun can be described by
\begin{eqnarray}\label{eq:dotN}
	\dot{N} = C - C_a N^2 - C_E N.
\end{eqnarray}
Here,
\begin{eqnarray}
	\frac{\hbox{d}C}{\text{d}^3 \mathbf{r} \text{d}^3\mathbf{v} \text{d}^3 \mathbf{v}_A\text{d}\Omega} &=& \sum_A f_A(\mathbf{r},\mathbf{v}_A) f(\mathbf{r},\mathbf{v}) \label{eq:cap}\\
	& & \times \: |\mathbf{v} - \mathbf{v}_A|\frac{\text{d}\sigma_A}{\text{d}\Omega} \: \Bigg|_{v_f < v_{esc}}, \nonumber
\end{eqnarray}
is the capture rate of WIMPs in the Sun by elastic scattering of halo WIMPs with distribution function $f$ off solar nuclei with atomic number $A$ and distribution function $f_A$. The interaction cross section is $\hbox{d}\sigma_A/\hbox{d}\Omega$.  WIMPs are only considered captured if the post-scatter speed of the WIMP $v_f$ is less than the local escape speed $v_{esc}(r)$.  The coefficient in the second term of Eq. (\ref{eq:dotN}), $C_a$, is defined by
\begin{eqnarray}\label{eq:gamma_N}
	\Gamma_a  = \frac{1}{2} C_a N^2.
\end{eqnarray}
If the number density can be described by Eq. (\ref{eq:num_sep}), $C_a$ is constant and has the form
\begin{eqnarray}
	C_a = 2\langle \sigma v \rangle_a \int \text{d}^3\mathbf{r} \, \tilde{n}^2(r).
\end{eqnarray}
If the number density is not in the form (\ref{eq:num_sep}), then $C_a$ will be time-dependent and it will also be necessary to model the time-evolution of the density profile.  The annihilation term in Eq. (\ref{eq:dotN}) is twice the annihilation rate because most theoretically motivated WIMPs are self-annihilating; each annihilation event removes two WIMPs from the Sun.  The last term in Eq. (\ref{eq:dotN}), $C_E N$, is an evaporation rate.  This term is important for very low mass WIMPs ($m_\chi \lesssim 4$ GeV), since light WIMPs may gain enough energy from interactions with nucleons in the Sun to become unbound to the solar system \cite{gaisser1986}, but is negligible for the range of WIMP masses ($m_\chi > 40$ GeV) generically expected in extensions to the SM.

If $C$ and $C_a$ are time-independent and $C_E$ is negligible, Eq. (\ref{eq:dotN}) has the solution
\begin{eqnarray}\label{eq:n_t}
	N(t) = \sqrt{C/C_a} \tanh(t/t_e),
\end{eqnarray}
where 
\begin{eqnarray}
	t_e = 1/\sqrt{CC_a}\label{eq:t_e}
\end{eqnarray}
is the timescale for the number of WIMPs in the Sun to reach equilibrium. Eq. (\ref{eq:n_t}) is subject to the boundary condition that $N(0) = 0$ at the birth of the solar system.  The annihilation rate goes as
\begin{eqnarray}
	\Gamma_a = \frac{1}{2}C \tanh^2(t/t_e), \label{eq:gamma_tanh}
\end{eqnarray}
which has the limits
\begin{eqnarray}
	\Gamma_a (t) = \begin{cases}
		\frac{1}{2} C &\text{if } t/t_e \gg 1 \\
		\frac{1}{2} C (t/t_e)^2 = \frac{1}{2}C^2 C_a t^2 &\text{if } t / t_e \ll 1 \label{eq:gamma_c}
		 
	\end{cases}
\end{eqnarray}

In the instantaneous thermalization model, the WIMPs captured in the Sun settle quickly to an equilibrium (i.e., the thermalization time $t_t \ll t_e$). Thus, the WIMP density profile is separable in the sense of Eq. (\ref{eq:num_sep}) and can be described by 
\begin{eqnarray}\label{eq:num_dens}
	n_e(r,t) = n_c (t) e^{-m_\chi \Phi(r)/k_B T},
\end{eqnarray}
where $n_c (t)$ is determined by $N(t)$, $m_\chi$ is the WIMP mass, $\Phi(r)$ is the gravitational potential of the Sun, and $T$ is the characteristic temperature of WIMPs in the Sun (typically, the core temperature of the Sun) \cite{griest1987,gould1987b,edsjo1995,bahcall2005}.  Since the density profile is fixed as a function of time, the solution Eq. (\ref{eq:n_t}) applies, and we can express the timescale $t_e$ in units of the age of the solar system $t_\odot = 4.5 \hbox{ Gyr}$,
\begin{eqnarray}
	\frac{t_\odot}{t_e} &\approx& 74 \left[ \frac{C}{10^{30} \hbox{ yr}^{-1} } \right]^{1/2} \left[ \frac{\langle \sigma v \rangle_a}{3\times 10^{-26} \hbox{ cm}^3 \hbox{ s}^{-1} } \right]^{1/2} \\
	& &\times \: \left[ \frac{ m_\chi}{100\hbox{ GeV}} \right]^{3/4}. \nonumber
\end{eqnarray}
Here, $\langle \sigma v \rangle_a = 3\times 10^{-26} \hbox{ cm}^3 \hbox{ s}^{-1}$ is the annihilation cross section necessary to create thermal relic WIMPs at the abundance observed in the universe \cite{jungman1996}.  $C = 10^{30} \hbox{ yr}^{-1}$ is the typical capture rate in the Sun by hydrogen for a WIMP with $m_\chi \approx 100$ GeV and WIMP-proton elastic scattering cross section $\sigma_p = 10^{-40} \hbox{ cm}^2$ assuming that the mass density of WIMPs in the Galactic halo is $\rho_\chi = 0.3 \hbox { GeV cm}^{-3}$ and that the local WIMP population is smooth and approximately described by a Maxwellian velocity distribution (with one-dimensional velocity dispersion $\sigma = v_\odot / \sqrt{2}$, where $v_\odot$ is the speed of the Local Standard of Rest).  If $m_\chi \gg m_A$, where $m_A$ is the mass of the target nucleus, the capture rate goes as
\begin{eqnarray}\label{eq:capture_scale}
	C \propto \rho_\chi m^{-2}_\chi \sigma_A,
\end{eqnarray}
where $\sigma_A$ is the elastic scattering cross section.  The dependence of the capture rate on the WIMP mass reflects not only the factor of $m^{-1}_\chi$ from the WIMP number density $n_\chi = \rho_\chi / m_\chi$ for fixed WIMP mass density but also kinematic suppression since it is difficult for halo WIMPs to transfer enough energy to solar nuclei to become bound to the Sun.  More detailed calculations of the capture rate can be found, e.g., in \cite{gould1992}.  For a WIMP mass of $m_\chi = 100$ GeV, the number of WIMPs in the Sun will have reached equilibrium unless \sigmapsi$\lesssim 10^{-45} \hbox{ cm}^2$ or \sigmapsd$\lesssim 10^{-43} \hbox{ cm}^2$.  Thus, the annihilation rate can be computed once the WIMP mass $m_\chi$, spin-dependent and spin-independent elastic scattering cross sections, the annihilation cross section, and the local halo WIMP phase space density are known.

\section{The Simulations}\label{sec:sim}
In Paper I, we described a set of simulations to determine the lifetimes and distribution function at the Earth of WIMPs bound to the solar system by elastic scattering off solar nuclei.  In order to understand how those quantities depend on WIMP mass and elastic scattering cross section, we ran four sets of simulations, each with a different combination of WIMP mass and elastic scattering cross section ($m_\chi = 60$ AMU, \sigmapsi$\: =10^{-41} \hbox{ cm}^2$; $m_\chi = 60 \hbox{ AMU}$, \sigmapsi$\: =10^{-43}$ cm$^2$; $m_\chi = 150$ AMU, \sigmapsi$\: =10^{-43}$ cm$^2$; $m_\chi = 500\hbox{ AMU}$, \sigmapsi$\: =10^{-43}$ cm$^2$).  In all cases, we set \sigmapsd$\: =0$ in order to simplify the interpretation of the results, but in Paper I, we describe how to extrapolate the results to regimes in which spin-dependent scattering dominates.  

We modeled the halo WIMPs as having a Maxwellian velocity distribution in the Galactocentric frame, setting the one-dimensional velocity dispersion to $\sigma = v_\odot / \sqrt{2}$.  We used the standard solar model described in \cite{bahcall2005} to model the Sun.  Scatters were treated in the ``cold Sun'' approximation, in which the thermal speeds of the solar nuclei are neglected.  This is a reasonable assumption since the WIMP kinetic energy $K_\chi \gg K_A$ for any nuclear species $A$, and the halo WIMP speed $v_\chi \gg v_A$.

Each simulation followed $10^5 - 10^6$ WIMPs from the initial time and location of the initial scatter by a solar nucleus of an unbound WIMP onto a bound orbit.  We followed the orbits using a symplectic integrator with an adaptive time step \cite{mikkola1999,preto1999}, with passages through the Sun, close encounters with the planets, and nearly unbound orbits treated as special cases.  The simulations allowed for the possibility of additional elastic scattering in the Sun whenever the orbits traversed the solar interior.  In order to more easily interpret the results of the simulation, we used a simplified solar system consisting only of the Sun and Jupiter.  Jupiter was placed on a circular orbit about the Sun.  This restricted three-body problem admits a constant of motion, the Jacobi constant, which is a useful check on the accuracy of the integration code.  Typically, errors in the Jacobi constant were oscillatory and no more than a few parts in $10^7$ at aphelion.  The details of the integration scheme are given in Paper I.

The integrations were terminated if the orbits met one of three conditions: (i) The WIMP rescattered onto an orbit that was no longer Earth-crossing. Such orbits thermalize quickly in the Sun, as discussed below, and hence were no longer relevant to our study of the WIMP distribution function at the Earth.  (ii) The particle was ejected from the solar system.  (iii) The WIMP survived on an Earth-crossing orbit for the entire lifetime of the solar system, $t_\odot = 4.5\hbox{ Gyr}$.  We use the lifetime distributions as a function of the semi-major axis of the WIMP orbits to construct our argument below.

\section{Insights from WIMP Orbit Simulations}\label{sec:insight}

The simulations offer insight into two possible mechanisms that may affect the annihilation rate in the Sun: the finite time required to thermalize WIMPs in the Sun, and changes to WIMP trajectories due to gravitational interactions with bodies in the solar system other than the Sun.  The former would be relevant even if the Sun were an isolated body, while the latter depends on the presence of planets in the system.  We show that both suppress the annihilation rate by an amount that increases with increasing WIMP mass, and the suppression is stronger for spin-dependent than spin-independent interactions at fixed WIMP mass.  

\subsection{Rescattering Times}\label{subsec:rescatt}
In the standard picture of WIMP annihilation, it is assumed that WIMPs thermalize with solar nuclei on very short timescales.  In the absence of planets, the time for a particle to rescatter after it becomes bound to the solar system is
\begin{eqnarray}
	t_r \sim P_\chi / \tau, \label{eq:rescatt}
\end{eqnarray}
where $P_\chi$ is the WIMP orbital period and $\tau$ is the optical depth for one passage through the Sun.  The thermalization time is $t_t = Xt_r$, where the factor $X$, which we will estimate later in this section, depends on the WIMP mass.  If $t_t$ is longer than the age of the solar system $t_\odot$, the particles do not thermalize in the Sun, and the annihilation rate of WIMPs in the Sun essentially vanishes.  To find the regions in WIMP parameter space for which $t_t \gtrsim t_\odot$, we consider how the thermalization time depends on both the WIMP cross section and mass. 

The thermalization time depends on the WIMP-bar\-yon cross section through the optical depth $\tau$, which is proportional to the cross section $\sigma$.  In our simulations $\sigma_p^{SI} = 10^{-43} \hbox{ cm}^2$ corresponds to $\tau \sim 10^{-5}$, the exact value of which depends on the trajectory of the WIMP through the Sun.  We find that for a fixed cross section, $\tau_{SI} \sim 100 \, \tau_{SD}$ (the exact value depends weakly on mass, for reasons described in Paper I), so that it takes a much higher spin-dependent cross section to reach an equivalent optical depth as a spin-independent cross section.  This is because (i) hydrogen is the only species in the Sun with an appreciable spin-dependent capture probability, and (ii) even though there are only trace amounts of metals (with nucleon number $A > 4$) in the Sun ($\sim 2\%$ of the total mass), the spin-independent cross section per nucleus goes as $\sigma^{SI} \propto A^4$ (this is only moderately suppressed due to incoherent scattering).  Therefore, for a fixed WIMP-proton cross section, the rescattering time is much longer if spin-dependent interactions dominate in the Sun rather than spin-independent interactions.

The WIMP thermalization time depends on $m_\chi$ through several different effects.  As $m_\chi$ increases, the momentum transfer in an elastic scattering event corresponds to a smaller velocity change in the WIMP.  It thus becomes more difficult to scatter halo WIMPs onto bound orbits.  Those that are scattered onto bound orbits will have larger (less negative) orbital energy and hence longer orbital periods.  In the case of spin-dependent scattering, the median semi-major axis for $m_\chi = 100\hbox{ GeV}$ is $a \approx 0.05\hbox{ AU} \approx 10\, R_\odot$ (where $R_\odot$ is the radius of the Sun and $1 \hbox{ AU} \approx 215 \, R_\odot$), $a\approx 0.5\hbox{ AU}$ for $m_\chi = 1\hbox{ TeV}$, and $a \approx 5 \hbox{ AU}$ for $m_\chi = 10\hbox{ TeV}$.  If spin-independent scattering dominates in the Sun, the median scattered semi-major axis is $a\approx R_\odot$ for $m_\chi = 100\hbox{ GeV}$, $a\approx 0.05\hbox{ AU}$ for $m_\chi = 1\hbox{ TeV}$, and $a\approx 0.5\hbox{ AU}$ for $m_\chi = 10\hbox{ TeV}$.  The median semi-major axis is well approximated by
\begin{eqnarray}\label{eq:a_m}
	a_m \approx \frac{m_\chi}{m_A}\frac{R_\odot}{10} 
\end{eqnarray}
for both spin-independent and spin-dependent scattering.  Typically, $m_A \sim 10\hbox{ GeV}$ for spin-independent scattering and $m_A \sim 1 \hbox{ GeV}$ for spin-dependent scattering.  Since Kepler's third law states that $P_\chi \propto a^{3/2}$, we find that the median WIMP orbital period goes as $P_\chi \propto m_\chi^{3/2}$.

Once the WIMP becomes bound to the solar system, the number of scatters required to thermalize a WIMP increases with increasing mass, hence increasing $X$, the ratio of the thermalization time to the rescattering time.  To estimate $X$, we consider the typical amount of energy lost in each scatter.  In an inertial frame moving with the target nucleus (in the cold Sun approximation, this is just the heliocentric frame), the WIMP loses an energy
\begin{eqnarray}
	Q = \frac{\mu^2_A}{m_A} v^2 (1-\cos \theta)
\end{eqnarray}
per scatter, where $\mu_A = m_A m_\chi / (m_A + m_\chi)$ is the reduced mass of the WIMP-nucleus system, $v$ is the speed of the WIMP relative to the nucleus, and $\cos\theta$ is the center-of-mass scattering angle.  For a high mass WIMP, $\mu_A \sim m_A$, so the typical WIMP energy loss is
\begin{eqnarray}
	Q \sim m_A v_{esc}^2,
\end{eqnarray}
where $v_{esc}$ is the local escape speed from the Sun ($v_{esc} = 618 \hbox{ km s}^{-1}$ at the surface of the Sun, and is $\sim 2\times$ that value at the center of the Sun).  For now, we approximate scatters as occurring at points in the Sun where the escape velocity is not too different from that at the surface.  One can express the energy loss in terms of a semi-major axis, using
\begin{eqnarray}
	\frac{GM_\odot m_\chi}{2a_q} &\equiv& Q \\
		&\sim & m_A v_{esc}^2. \label{eq:aqapprox}
\end{eqnarray}
Solving for $a_q$, we find
\begin{eqnarray}
	a_q &\sim & \frac{GM_\odot}{2v_{esc}^2} \frac{m_\chi}{m_A} \\
	    &\sim & \frac{m_\chi}{m_A} R_\odot,
\end{eqnarray}
where we have used the fact that $v_{esc}^2 = 2 GM_\odot / R_\odot \sim G M_\odot / R_\odot$.  If a WIMP had a semi-major axis $a_i$ prior to scatter, its post-scatter semi-major axis $a_f$ can be described by
\begin{eqnarray}
	-\frac{GM_\odot m_\chi}{2a_f} &=& -\frac{GM_\odot m_\chi}{2a_i} - Q.	
\end{eqnarray}
We find
\begin{eqnarray}
	a_f &\sim& \frac{a_i a_q}{a_i + a_q},
\end{eqnarray}
and hence the change in semi-major axis per scatter is
\begin{eqnarray}
	\Delta a \equiv a_f - a_i = - \frac{a_i^2}{a_i + a_q}.
\end{eqnarray}

We can define a differential equation for the shrinking of the semi-major axis with time, 
\begin{eqnarray}
	\frac{\hbox{d}a}{\hbox{d}t} = - \tau \frac{a^{1/2}}{a + a_q},
\end{eqnarray}
where we have used $t = P_\chi / \tau$ as the time between scatters, and where we have expressed $a$ in AU and $t$ in years.  This equation has the solution
\begin{eqnarray}
	t_f = \frac{2}{\tau} \left[ \frac{1}{3} \left( a_i^{3/2} - a_f^{3/2} \right) + a_q \left(a_i^{1/2} - a_f^{1/2}\right) \right]\hbox{ yr},\label{eq:tf}
\end{eqnarray}
where again, all semi-major axes should be in units of AU.

\begin{figure}
	\includegraphics[width=3.3in]{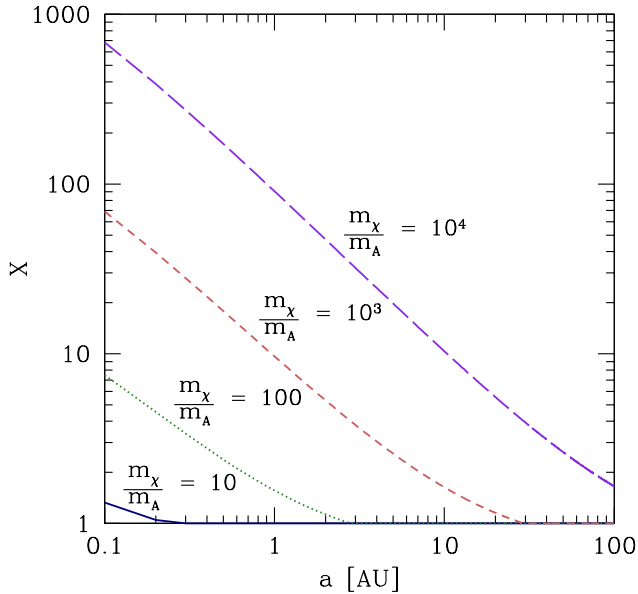}
	\caption{\label{fig:xfactor}The number of scatters $X$ required to bring a WIMP with initial semi-major axis $a$ down to $a_f = 2 R_\odot$ as a function of the WIMP-nucleus mass ratio.}
\end{figure}

To estimate the $X$ coefficient, we find $X = t_f \tau / P_\chi (a_i)$ using the time $t_f$ it takes for a WIMP with initial semi-major axis $a_i$ to reach a semi-major axis $a_f = 2R_\odot$, since the thermalization time $t_t$ is dominated by the scatters that occur when the orbital period is still fairly large.  We use $a_f = 2R_\odot$ because we consider scatters that occur near $R_\odot$, and because for semi-major axes much smaller than this, the thermal speeds of the nuclei (which we ignore here) become important.  We restrict $X$ to be strictly greater than unity, since an integer number of scatters is required to thermalize a WIMP.  Setting $a_f$ a factor of ten higher changes $X$ by only a few percent.  We show $X$ as a function of $a$ and $m_\chi / m_A$ in Fig. \ref{fig:xfactor}.  From this figure, we see that for a WIMP orbit with $a_i = 1\hbox{ AU}$, it takes $\mathcal{O}(1)$ scatter to thermalize if $m_\chi / m_A = 100$, and that $X\propto m_\chi / m_A$ for higher values of the mass ratio.

We can also use Eq. (\ref{eq:tf}) to estimate the lower limit on the elastic scattering cross section for which the median thermalization time of captured WIMPs $t_t$ is less than the age of the solar system $t_\odot$ as a function of $m_\chi / m_A$.  Using Eq. (\ref{eq:a_m}) to describe the median semi-major axis of captured WIMPs, we estimate the limits to be
\begin{multline}
	\sigma_p^{SI} \Big|_{\text{lim}} \approx  10^{-51} \hbox{ cm}^2 \Bigg\{ 0.3 \left( \frac{m_\chi}{m_A} \right)^{3/2} -1.4 \frac{m_\chi}{m_A} \\
	- 0.9 \Bigg\},
\end{multline}
and
\begin{eqnarray}
	\sigma_p^{SD} \Big|_{\text{lim}} \approx 100 \; \sigma_p^{SI} \Big|_{\text{lim}}. 
\end{eqnarray}
If spin-independent scattering dominates the capture rate in the Sun, the two solar species that are responsible for most of the capture rate are helium and oxygen, so $m_A \sim 10\hbox{ GeV}$.  Thus, the thermalization time exceeds the age of the solar system if $\sigma_p^{SI} \lesssim 10^{-51}\hbox{ cm}^2$ if $m_\chi = 100\hbox{ GeV}$, $\sigma_p^{SI} \lesssim 10^{-50}\hbox{ cm}^2$ if $m_\chi = 1\hbox{ TeV}$, and $\sigma_p^{SI} \lesssim 10^{-47}\hbox{ cm}^2$ if $m_\chi = 10\hbox{ TeV}$.  If spin-dependent scattering dominates in the Sun, the solar species that dominates the capture rate is hydrogen ($m_A \sim 1\hbox{ GeV}$), so the median thermalization time exceeds $t_\odot$ if $\sigma_p^{SD} \lesssim 10^{-48}\hbox{ cm}^2$ if $m_\chi = 100\hbox{ GeV}$, $\sigma_p^{SD} \lesssim 10^{-45}\hbox{ cm}^2$ if $m_\chi = 1\hbox{ TeV}$, and $\sigma_p^{SD} \lesssim 10^{-44}\hbox{ cm}^2$ if $m_\chi = 10\hbox{ TeV}$.

\subsection{The Effect of Jupiter}\label{subsec:planet}
We now investigate how the presence of Jupiter alters the WIMP lifetimes as a function of the initial semi-major axes.  In our simulations, the vast majority ($\sim 99.9\%$) of WIMPs that were initially scattered onto orbits with semi-major axis $a < 1.5\hbox{ AU}$ rescattered on timescales of order $t \approx P_\chi /\tau$.  It usually only required one scatter to reduce the semi-major axis below the threshold of interest because we considered spin-independent interactions only and simulated WIMPs with $m_\chi \leq 500 \hbox{ GeV}$.  For a given WIMP mass and scattering cross section, the thermalization time for these WIMPs with $a < 1.5\hbox{ AU}$ is described by Section \ref{subsec:rescatt}.  For completeness, we note that there is also a long-lived population of WIMPs with $a < 1.5\hbox{ AU}$ on secular resonances (ones which pull the perihelia out of the Sun for extended periods of time) that contribute substantially to the bound WIMP distribution function at the Earth, but make up only a tiny fraction $\sim 10^{-3}$ of the total population of $0.5 \hbox{ AU} < a < 1.5\hbox{ AU}$ WIMPs.  These WIMPs contribute negligibly to the annihilation rate.

There are two populations of WIMPs captured onto bound orbits that have their lifetimes altered by gravitational torques from Jupiter.  WIMPs with $1.5\hbox{ AU} < a < 2.6 \hbox{ AU}$ (these WIMPs do not cross Jupiter's orbit; the largest possible aphelion for an orbit with $a < 2.6\hbox{ AU}$ is $2\times 2.6 \hbox{ AU} = 5.2 \hbox{ AU}$, which is Jupiter's semi-major axis), hereafter called the ``long-lived'' population, have rescattering times of order $t_r\sim 100 P_\chi / \tau$.  Through a combination of mean-motion and secular resonances, Jupiter pulls the perihelia of such WIMPs out of the Sun for a significant fraction of their lifetimes in the solar system.  This reduces the probability of rescattering for any given orbital period, hence increasing the rescattering time by about two orders of magnitude.

The thermalization time of this population of long-lived particles, $t_{ll}$, is dominated by the time required for a WIMP to rescatter to a semi-major axis $a < 1.5 \hbox{ AU}$.  To demonstrate this, we define $X_{1.5}$ to be Eq. (\ref{eq:tf}) multiplied by $\tau / P_\chi$ for $a_i = 2\hbox{ AU}$ (a typical semi-major axis for long-lived WIMPs) and $a_f = 1.5\hbox{ AU}$.  If $m_\chi / m_A = 10^3$, $X_{1.5} = 1$, and so the time required for a WIMP to drop below $a = 1.5\hbox{ AU}$ is $t_{1.5} = 100 X_{1.5} P_\chi / \tau = 100 P_\chi / \tau$.  Once the WIMP reaches $a = 1.5$, according to Fig. \ref{fig:xfactor}, it takes $t_t \sim 7 \, P_\chi (1.5\hbox{ AU}) / \tau$ to thermalize, which is much less than $t_{1.5}$.

WIMPs that initially scatter onto Jupiter-crossing orbits (``Jupiter-crossing population''; $a > 2.6\hbox{ AU}$) are ejected from the solar system on Myr timescales unless the optical depth in the Sun so high that the rescattering timescale $P_\chi/\tau$ is less than the timescale $t_{\text{\jupiter}}$ for torques from Jupiter to pull the orbital perihelia out of the Sun.  In our simulation with $m_\chi = 60 \hbox{ AMU}$, \sigmapsi$\: =10^{-41}\hbox{ cm}^2$, and \sigmapsd$\: =0$, we found that $\sim 75\%$ of all Jupiter-crossing WIMPs were ejected before rescattering in the Sun.  The equivalent spin-dependent cross section, if such interactions dominate in the Sun, is \sigmapsd$\: \sim 10^{-39} \hbox{ cm}^2$.  For the simulations with \sigmapsi$\: =10^{-43}\hbox{ cm}^2$ (equivalent to \sigmapsd$\,\sim 10^{-41} \hbox{ cm}^2$), the percentage of Jupiter-crossing orbits that are ejected increased to $>98\%$.  

For $m_\chi / m_A \lesssim 100$, it takes on average only $X\sim 1$ times the rescattering time to bring a Jupiter-crossing orbit down to $a \sim R_\odot$.  Therefore, for such mass ratios, Jupiter-crossing WIMPs will thermalize if \sigmapsi$\,\gtrsim 10^{-41} \hbox{ cm}^2$ or \sigmapsd$\,\gtrsim10^{-39} \hbox{ cm}^2$.  If $m_\chi / m_A$ is much higher, a Jupiter-crossing WIMP is not guaranteed to thermalize even if it does rescatter; it may rescatter onto another Jupiter-crossing orbit or become a member of the long-lifetime population.  We define $X_{2.6} = t_f \tau / P_\chi$ for $a_i = 4\hbox{ AU}$ and $a_f = 2.6\hbox{ AU}$ using Eq. (\ref{eq:tf}).  For a mass ratio of $m_\chi / m_A = 10^4$, $X_{2.6} \sim 5$, and $X_{1.5}\approx 9$ for the WIMP to go from $a = 2.6 \hbox{ AU}$ to $a = 1.5\hbox{ AU}$.  Therefore, for high mass ratios ($m_\chi / m_A \gtrsim 10^3$), Jupiter-crossing WIMPs will only thermalize if the elastic scattering cross section is significantly higher than \sigmapsi$\,\sim 10^{-41}\hbox{ cm}^2$ or \sigmapsd$\,\sim 10^{-39}\hbox{ cm}^2$.

\subsection{Mapping the Suppression in Parameter Space}\label{subsec:map}

The total suppression of the annihilation rate will depend on both the WIMP mass and the WIMP-baryon cross section, and on whether the scattering is spin-dependent or spin-independent.  To quantify the suppression, one must determine what fraction of captured WIMPs belong to each population, and solve a differential equation for the number density of WIMPs in the Sun given realistic thermalization times.

\begin{figure*}
	\includegraphics[width=3.3in,angle=270]{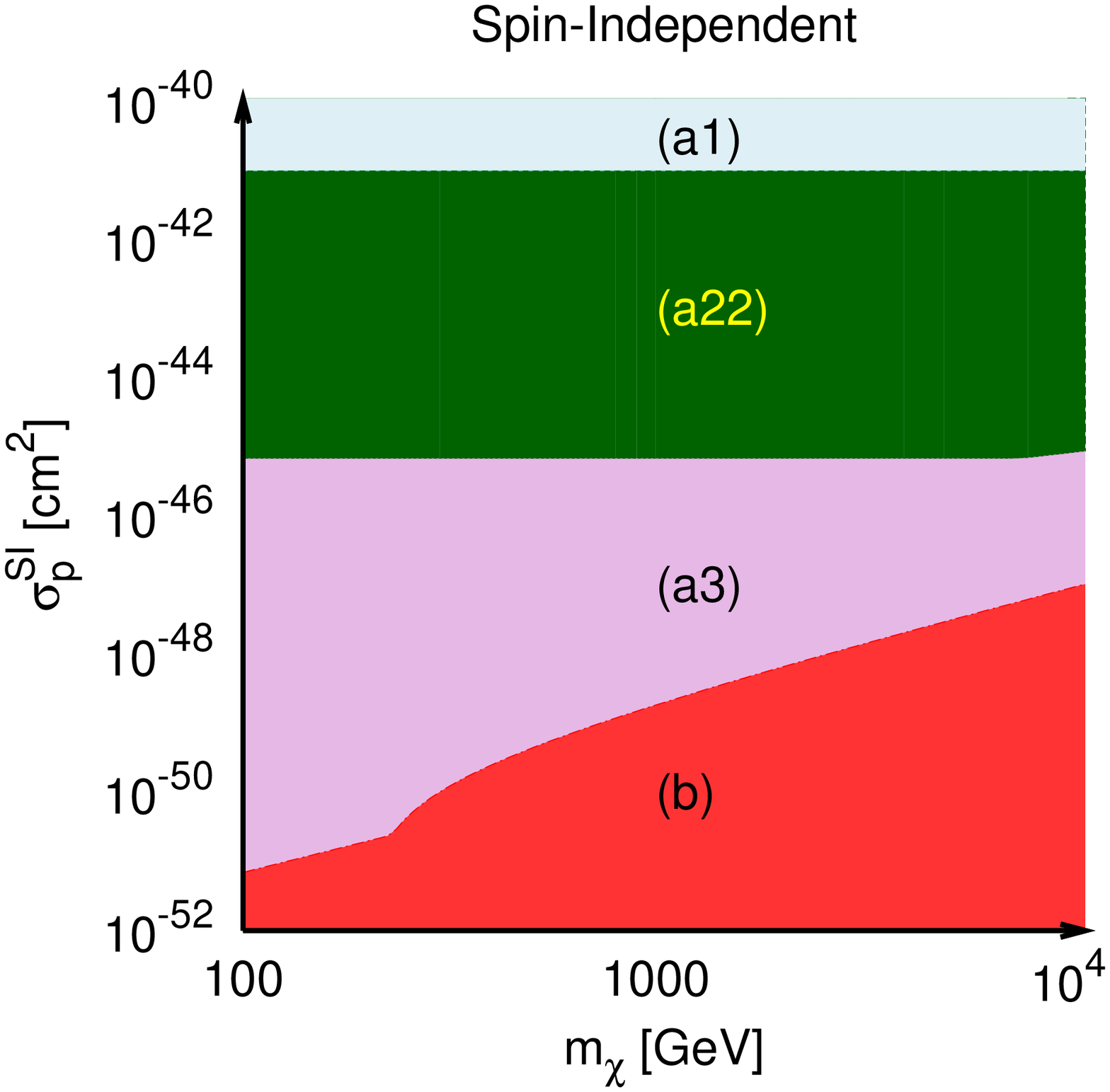} \includegraphics[width=3.3in,angle=270]{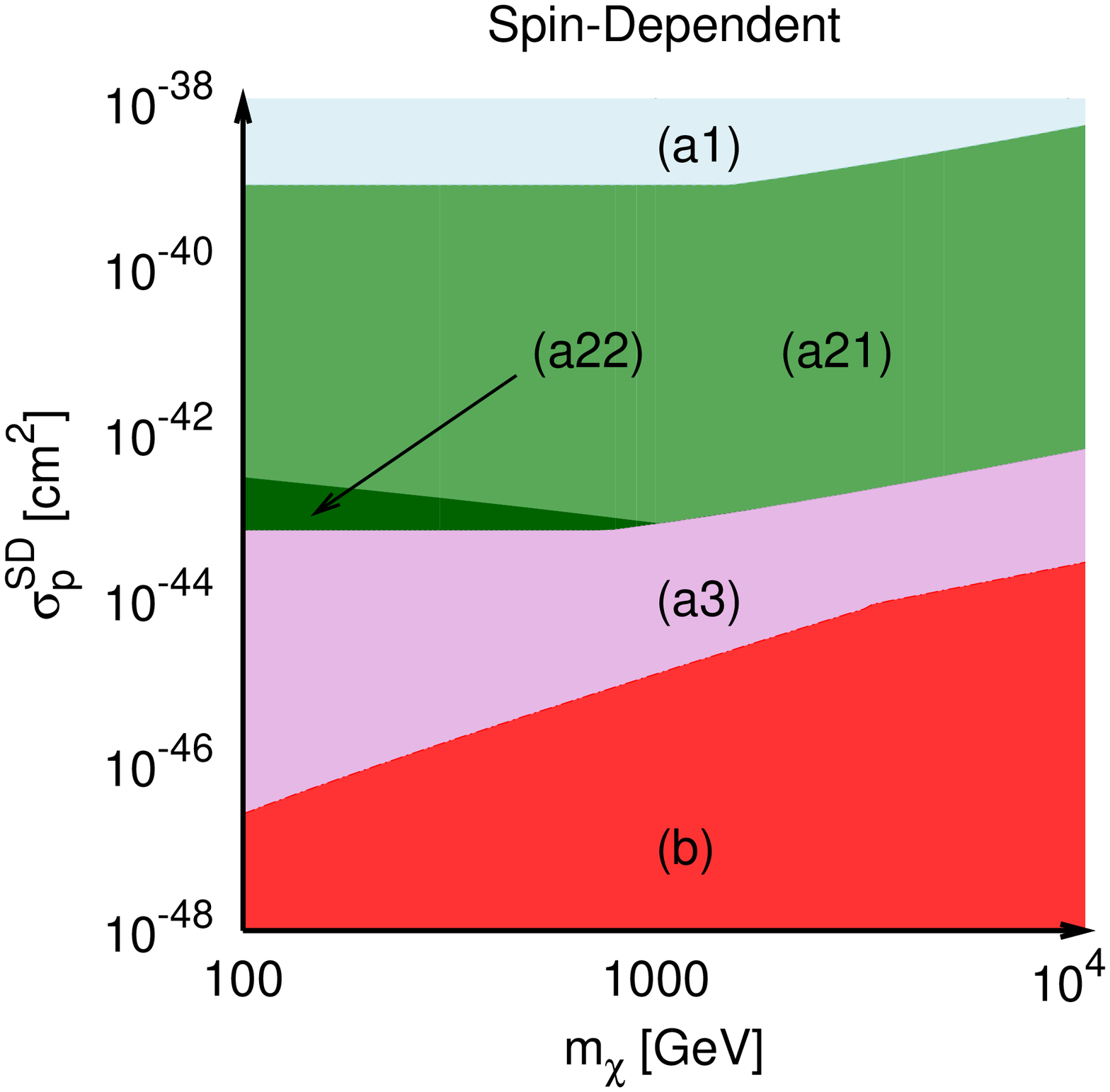}
	\caption{\label{fig:param}Suppression regimes, as defined in the text, as a function of WIMP mass and (\emph{left}) spin-independent and (\emph{right}) spin-dependent elastic scattering cross section.}
\end{figure*}

We classify the types of suppression of the annihilation rate in Fig. \ref{fig:param} according to the description below, and demonstrate how to calculate the number of WIMPs in the Sun (and hence, the suppression with respect to the instantaneous thermalization model) in each regime.  To simplify the discussion, we assume that each WIMP population can be described by its median thermalization time.  A more precise calculation of the annihilation rates would model the thermalization time distributions of each WIMP population.
  
(a) $X P_\chi / \tau < t_\odot$:  In this case, all WIMPs with $a < 1.5$ AU thermalize in the Sun.  We define the boundary in cross section for this part of parameter space by considering either the thermalization time for the median semi-major axis $a_m$ of all captured WIMPs if $a_m < 1.5 \hbox{ AU}$ or the thermalization time for $a = 1.5\hbox{ AU}$ if the median semi-major axis lies above this value.  For very small mass ratios ($m_\chi / m_A \lesssim 20$), $a_m$ is less than $a_f = 2R_\odot$, the semi-major axis we used in Section \ref{subsec:rescatt} to estimate the thermalization time.  For these values, we set $X=1$ to estimate the thermalization time.

This case breaks up into several sub-classes, depending on whether the Jupiter-crossing and long-lived WIMP populations thermalize in the Sun.
\begin{itemize}
	\item (a1) $t_\text{\jupiter} \gtrsim X_{2.6} P_\chi / \tau$: In this case, the timescale for Jupiter to pull the perihelia of Jupiter-crossing WIMPs out of the Sun $t^\text{\jupiter}$ is greater than the rescattering timescale, so Jupiter-crossing WIMPs thermalize in the Sun.  The thermalization time corresponding to $X_{2.6}$ is within a factor two for the range of $m_\chi$ in Fig. \ref{fig:param} if a more careful estimate using the median Jupiter-crossing semi-major axis is considered.  For the range of cross sections for which Jupiter-crossing WIMPs thermalize, long-lived WIMPs also thermalize on timescales of $t_{ll} = 100 X_{1.5} P_\chi / \tau$, and so the annihilation rate in the Sun will be unchanged from that computed in Section \ref{sec:std}.  

	\item (a2) $t_\text{\jupiter} \lesssim X_{2.6} P_\chi / \tau \hbox{ and } t_{ll} < t_\odot$:  In this case, Jupiter-crossing WIMPs will not thermalize, meaning that the capture rate of WIMPs that do thermalize is $C^{\prime} = C^{\text{\jupiter}} \equiv (1-f_{\text{\jupiter}})C$, where $C$ is the capture rate calculated in Eq. (\ref{eq:cap}) and $f_{\text{\jupiter}}$ is the fraction of all WIMPs on Jupiter-crossing orbits.  The capture rate may be further reduced depending the thermalization time of the long-lived WIMPs.  
	
	(a21) $t_{ll} < t^{\text{\jupiter}}_e$:  If the long-lived WIMPs thermalize quickly relative to $t^{\text{\jupiter}}_e$, the equilibrium time due to the reduced capture rate, then the annihilation rate of WIMPs in the Sun can be calculated in the instantaneous thermalization model of Section \ref{sec:std}, replacing $C$ with $C^\text{\jupiter}$ in the calculation.  Therefore, the annihilation rate will have the same form as Eq. (\ref{eq:gamma_tanh}), replacing $C$ with $C^{\text{\jupiter}}$.

	(a22) $t_{ll} > t^{\text{\jupiter}}_e$:  One can think of the capture rate of WIMPs in the Sun as a step-function,
	\begin{eqnarray}
		C^{\prime}(t) = \begin{cases} (1-f_{ll})C, &t < t_{ll} \\
				C^{\text{\jupiter}}, &t > t_{ll},
			\end{cases}
	\end{eqnarray}
	where $f_{ll}$ is the fraction of all WIMPs initially scattered onto bound orbits that have $a > 1.5 \hbox{ AU}$, and the capture rate reflects only those particles that may have thermalized on timescales $< t$.  Solving the differential equation (Eq. \ref{eq:dotN}) for the number of WIMPs in the Sun using this time-dependent capture rate, we find
	\begin{multline}
		N = \sqrt{\frac{C^\text{\jupiter}}{C_a}} \tanh\Bigg[ (t_\odot - t_{ll}) / t^{\text{\jupiter}}_e \\
		   + \: \tanh^{-1}\left( \sqrt{\frac{C_a}{C^\text{\jupiter}}} N_0 \right) \Bigg],
	\end{multline}
	with
	\begin{eqnarray}
		N_0 = \sqrt{\frac{(1-f_{ll})C}{C_a}} \tanh\left( \sqrt{(1-f_{ll})CC_a}t_{ll} \right).
	\end{eqnarray}

	\item (a3) $t_\text{\jupiter} \lesssim X_{2.6} P_\chi / \tau \hbox{ and } t_{ll} > t_\odot$:  None of the particles with $a > 1.5$ AU will thermalize in the Sun.  In this case, the annihilation rate has the same form as that shown in Eq. (\ref{eq:gamma_tanh}) but with a reduced capture rate $C^{\prime} = ( 1 - f_{ll} ) C$.

\end{itemize}

(b) $X P_\chi / \tau > t_\odot$:  A WIMP captured onto an orbit with the median semi-major axis $a_m$ will not thermalize in the Sun. 

The types of suppression have a stronger mass dependence for the range of $m_\chi$ in Fig. \ref{fig:param} if spin-dependent scattering dominates in the Sun because the typical mass of the nucleus on which the WIMPs scatter is a factor of ten smaller than for spin-independent scattering.  The regions are also shifted up for spin-dependent interactions because $\tau$ is a factor of 100 smaller than for spin-independent interactions for a fixed WIMP-proton cross section.  For the swath of parameter space shown in Fig. \ref{fig:param}, spin-independent cross sections in case (a2) are always in sub-case (a22) because, for fixed $\tau$, the capture rate is higher for spin-independent cross sections than for spin-dependent cross sections. Spin-dependent captures are more kinematically suppressed.  Therefore, $t^\text{\jupiter}_e$ is systematically shorter for spin-independent interactions than for spin-dependent interactions.

\begin{figure*}
	\includegraphics[width=3.3in]{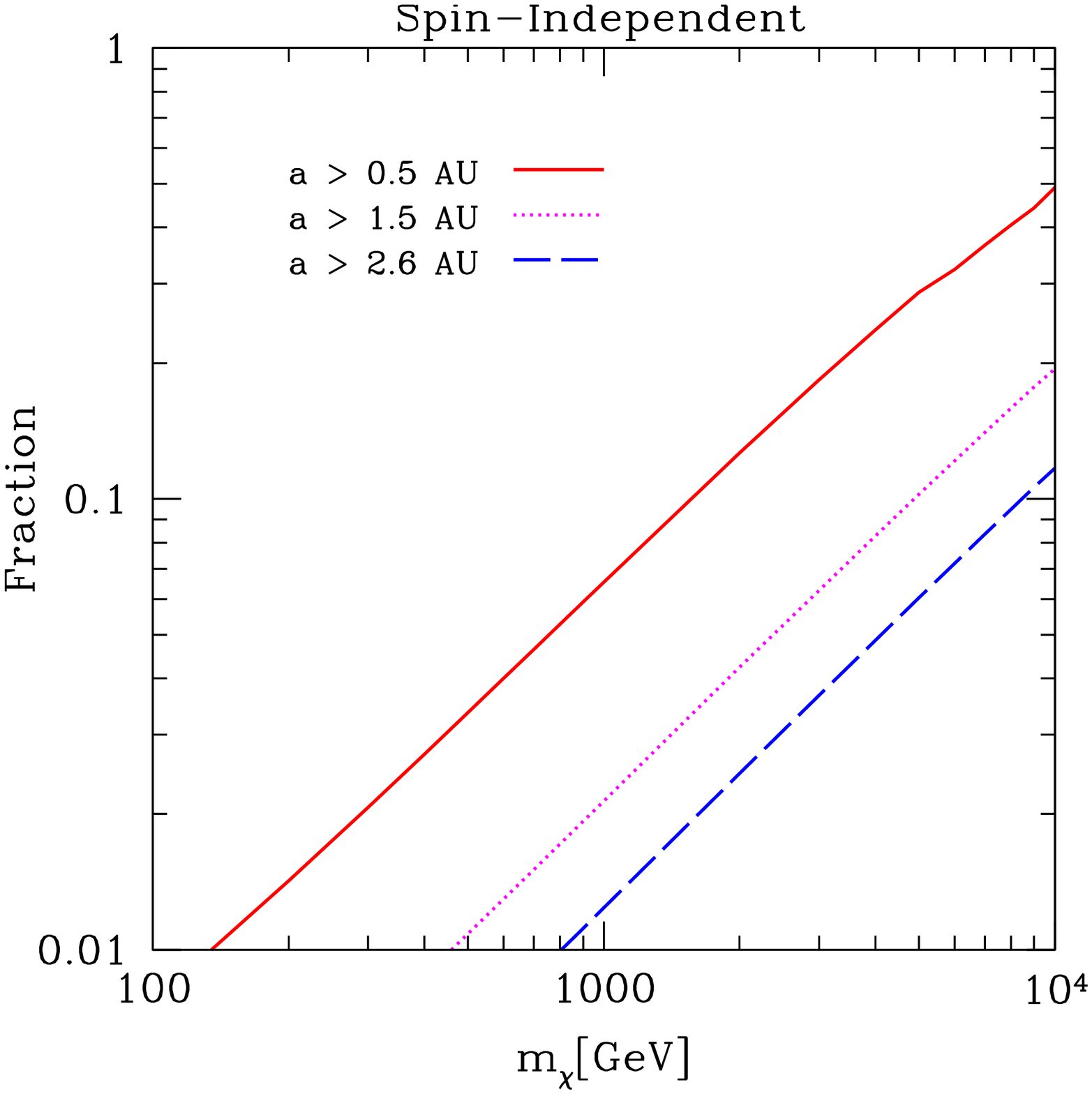}\includegraphics[width=3.3in]{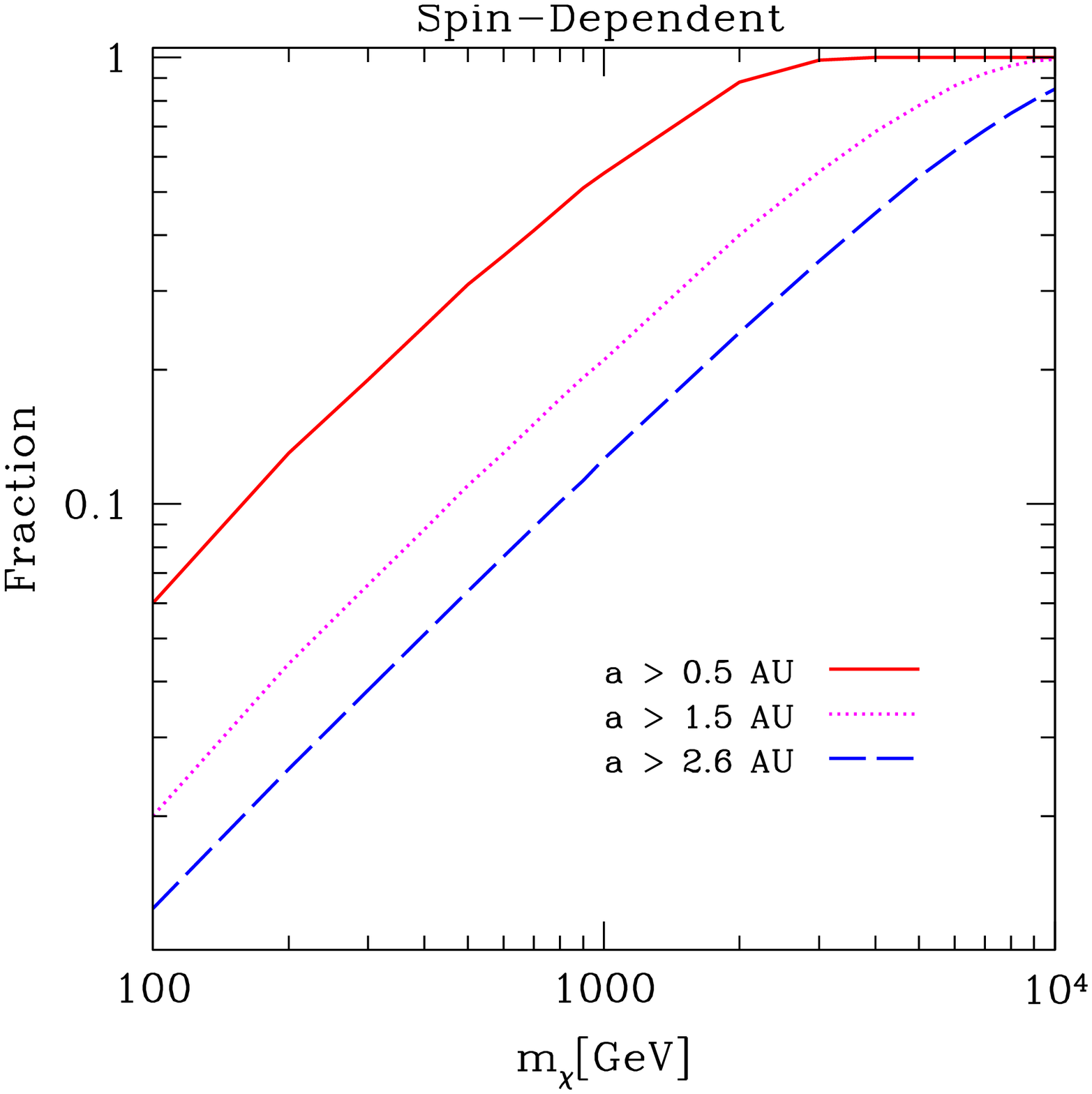}
	\caption{\label{fig:frac}The ratio of the capture rate of WIMPs onto orbits above a certain energy threshold to the total capture rate due to (\emph{left}) spin-independent and (\emph{right}) spin-dependent interactions in the Sun as a function of WIMP mass.}
\end{figure*}

We quantify the suppression as $\Gamma_a / \Gamma_a^0$, where $\Gamma_a^0$ is the annihilation rate in the instantaneous thermalization model, and $\Gamma_a$ is the annihilation rate calculated using the above methods.  For points in WIMP parameter space for which the gravitational torques from Jupiter suppress the annihilation rate, the suppression depends on $f_\text{\jupiter}$ and $f_{ll}$.  In Fig. \ref{fig:frac}, we show the those crossing fractions for the initial distribution of captured WIMPs for both spin-independent and spin-dependent interactions.  If spin-independent interactions dominate in the Sun, the fraction of captured WIMPs on Jupiter-crossing orbits is never high; it reaches only $f_\text{\jupiter} \sim 0.1$ if $m_\chi = 10\hbox{ TeV}$.  The fraction of WIMPs with $a> 1.5\hbox{ AU}$ is only $f_{ll} \sim 0.2$ for the same WIMP mass.  Therefore, we expect the suppression of the annihilation rate to be minimal is spin-independent interactions dominate in the Sun.

\begin{figure*}
	\includegraphics[width=3.3in]{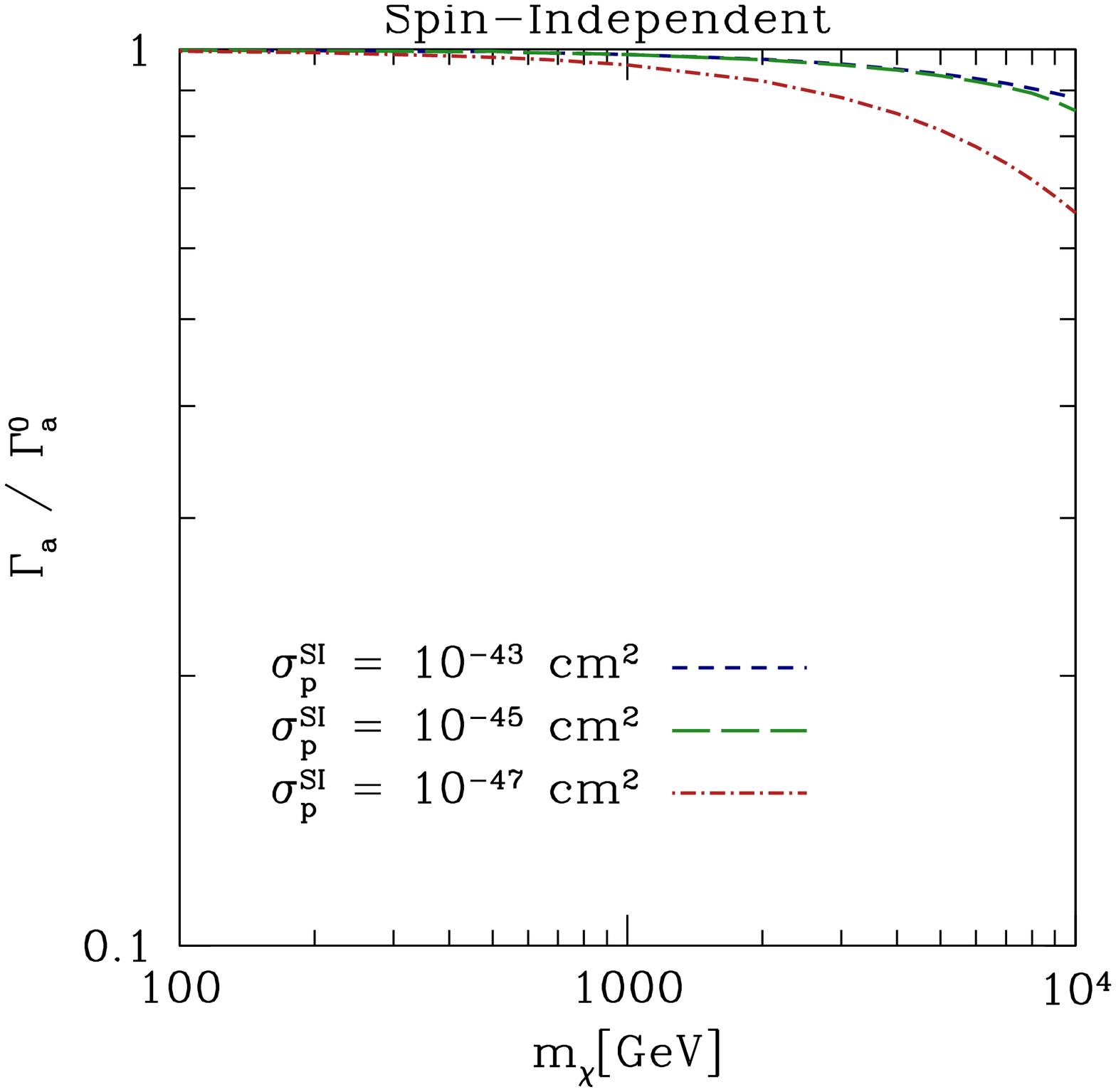} \includegraphics[width=3.3in]{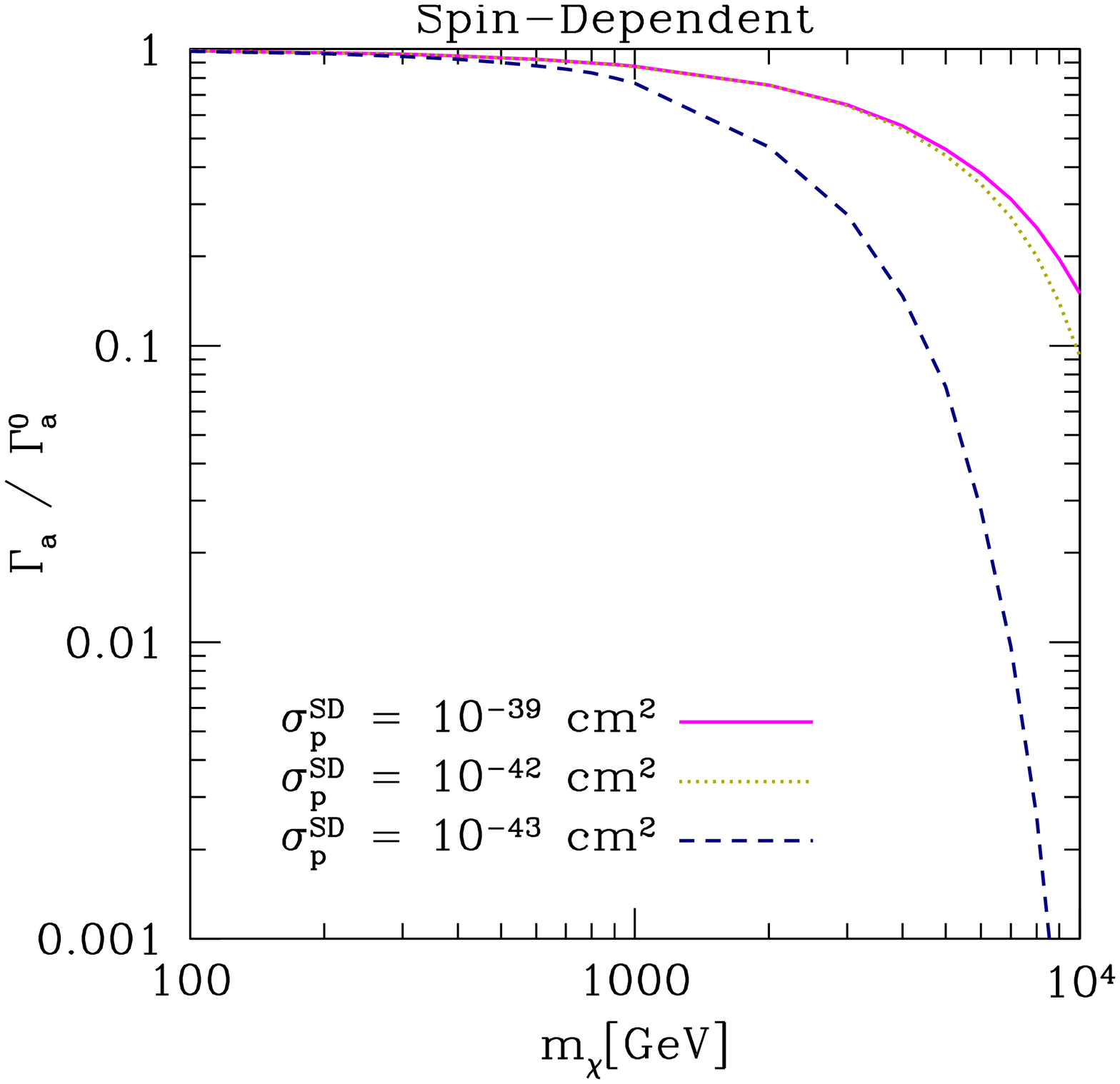}
	\caption{\label{fig:supp}The ratio of the estimated annihilation rate in the Sun to the annihilation rate calculated in the instantaneous thermalization model.  In all cases, $\langle \sigma v \rangle_a = 3\times 10^{-26} \hbox{ cm}^2$.}
\end{figure*}

In Fig. \ref{fig:supp}, we show the suppression as a function of WIMP mass for several cross sections.  If \sigmapsi$=10^{-43}\hbox{ cm}^2$, the annihilation rate is reduced by less than 10\% unless $m_\chi \gtrsim 10\hbox{ TeV}$.  Only Jupiter-crossing WIMPs fail to thermalize.  For such a cross section the number equilibrium time is short compared to the age of the solar system, so $\Gamma_a / \Gamma_a^0 \propto (1-f_{\text{\jupiter}})$ (Eq. \ref{eq:gamma_c}).  The situation is virtually unchanged if \sigmapsi drops by two orders of magnitude; the only change is that at $m_\chi = 10\hbox{ TeV}$, the annihilation rate drops below equilibrium.  If \sigmapsi$=10^{-47}\hbox{ cm}^2$, the suppression increases since the long-lived WIMPs have thermalization lifetimes beyond $t_\odot$ for all WIMP masses shown in Fig. \ref{fig:param}, and the number equilibrium time also exceeds $t_\odot$.  In this case, the annihilation rate is reduced by more than 10\% for $m_\chi \gtrsim 2\hbox{ TeV}$.

The suppression is far more pronounced if spin-dependent scattering dominates in the Sun.  According to Fig. \ref{fig:frac}, nearly all WIMPs have $a > 1.5\hbox{ AU}$ if $m_\chi = 10\hbox{ TeV}$, and 85\% of captured WIMPs are on Jupiter-crossing orbits.  Stepping through the lines in Fig. \ref{fig:supp} for spin-dependent scattering, we find that for \sigmapsd$\: =10^{-39} \hbox{ cm}^2$, the suppression is linear in the capture rate $C^\text{\jupiter}$ since $t_e \ll t_\odot$.  The annihilation rate is reduced by $>10\%$ from the instantaneous thermalization model for $m_\chi \gtrsim 1 \hbox{ TeV}$.  For \sigmapsd$= 10^{-42} \hbox{ cm}^2$, the suppression increases for $m_\chi \gtrsim 5\hbox{ TeV}$ since the number equilibrium timescale for the full capture rate is of order the age of the solar system.  Thus, the annihilation rate is reduced from its steady-state value.  The equilibrium timescale is increased for high masses since the total capture rate $C$ decreases with WIMP mass as $C\propto m_\chi^{-2}$ for high $m_\chi$.  

For \sigmapsd$\: =10^{-43} \hbox{ cm}^2$, long-lived WIMPs will have lifetimes exceeding the age of the solar system if $m_\chi \gtrsim 500\hbox{ GeV}$.  Moreover, the number equilibrium time of the WIMPs that do thermalize $t_e \gtrsim t_\odot$ so that the suppression $\Gamma_a / \Gamma_a^0 \propto (1-f_{ll})^2$ (see Eq. \ref{eq:gamma_c}).  The suppression for $m_\chi \sim 1\hbox{ TeV}$ is $\Gamma_a \approx 0.7 \, \Gamma_a^0$, and $\Gamma_a < 0.01\,\Gamma_a^0$ if $m_\chi \gtrsim 7\hbox{ TeV}$.  If the cross section is much lower than this, $\sim 10 \hbox{ TeV}$ WIMPs will not thermalize at all.

\section{Discussion}\label{sec:conclusion}
Throughout this work, we have considered a stripped-down solar system with Jupiter as the only planet.  We now ask how the above discussion is changed by the inclusion of the other planets in our solar system.  The presence of the other planets may affect the results above if they either (i) change the lifetimes for each population of WIMP ($a < 1.5 \hbox{ AU}$, $1.5 \hbox{ AU} < a < 2.6 \hbox{ AU}$, and $a > 2.6\hbox{ AU}$) or (ii) if they introduce new classes of WIMP populations.  

Given the following arguments, it is unlikely that the presence of other planets will affect either the lifetimes or the classification of WIMP orbits enough for the annihilation rate to be much different from those calculated in Section \ref{sec:insight}.  First, we consider Jupiter-crossing WIMPs, with $a > 2.6 \hbox{ AU}$.  Since Jupiter is by far the most massive planet in the solar system, it largely sets the ejection timescale of WIMPs from the solar system.  Thus, the population of $a > 2.6 \hbox{ AU}$ WIMPs should be largely unaffected by the presence of the other planets.

WIMPs with $a < 2.6 \hbox{ AU}$ could be affected if either the resonance structure is significantly altered (recall that the long-lived population owes its survival to both mean-motion and secular resonances) or if there is a significant probability that \emph{most} WIMPs of a given initial semi-major axis will experience a strong encounter with an inner planet.  The planets other than Jupiter are not likely to affect the secular resonance structure for the highly eccentric orbits that originate in the Sun.  Simulations of near-Earth asteroid orbits show that changes to the resonance structure due to the all non-Jupiter planets are only important if the orbits are initially circular, at low inclination, and have semi-major axes quite near those of either the Earth or Venus \cite{michel1996}.  However, WIMPs captured in the Sun will generically have very high eccentricities if they cross the orbits of the inner planets, so it is likely that Jupiter still dominates the resonance structure for most orbits.  

To estimate the importance of close planetary encounters, we treat interactions as local and describe changes in the WIMP semi-major axis using a random-walk approximation.  This treatment may not be a good description for the scattering of WIMPs on resonances, since such WIMPs may either be protected from \cite{michel1996} or have much stronger interactions with planets than predicted in the diffusion approximation.  However, this argument will supply some rough interaction timescales.  We consider inner planet encounters to have a significant impact on WIMP lifetimes only if the resulting RMS change to the WIMP semi-major axis $\langle (\delta a)^2 \rangle / a^2 \sim 1$.  

If we approximate the encounters to be local, the typical change to the WIMP speed $u$ in an inertial frame moving with planet $P$ is
\begin{eqnarray}
	\delta u \sim \frac{ GM_P}{b u}
\end{eqnarray}
for an impact parameter $b$.  Since the WIMP orbits are highly eccentric, the heliocentric velocity $\mathbf{v}$ of the WIMP is nearly orthogonal to the heliocentric velocity of the planet $\mathbf{v}_P$ (since the inner planets are on nearly circular orbits) unless the WIMP semi-major axis $a \approx a_P$, so that
\begin{eqnarray}
	u \approx \sqrt{ v^2 + v_P^2},
\end{eqnarray}
where $v_P^2 = GM_\odot / a_P$ is the square of the planet's orbital speed.  Therefore, in heliocentric coordinates, the change to the WIMP's speed is
\begin{eqnarray}
	\delta v \sim \frac{GM_P}{bv}.
\end{eqnarray}
The change in semi-major axis in each encounter is
\begin{eqnarray}
	\delta a &\sim& \frac{a^2}{GM_\odot} v\delta v \\
		&\sim& \frac{M_P}{M_\odot} \frac{a^2}{b} \label{eq:delta_a}
\end{eqnarray}
since the WIMP's energy $E = -GM_\odot / 2a = v^2/2 + \Phi_\odot(r)$.

The RMS change to the semi-major axis is
\begin{eqnarray}\label{eq:rms_a}
	\langle (\delta a)^2 \rangle = N (\delta a)^2,
\end{eqnarray}
where $N$ is the number of encounters with impact parameter $\leq b$.  To determine $N$, we estimate that a WIMP has a probability $\sim (b/a_P)^2$ (the solid angle subtended by a sphere of radius $b$ centered on the planet, as seen from the center of the Sun) of having an encounter with impact parameter $\leq b$ during each WIMP orbital period $P_\chi$.  Therefore, if we consider the WIMP for a time $t$, the total number of encounters in this time is
\begin{eqnarray}\label{eq:Nt}
	N \sim \frac{t}{P_\chi} \left( \frac{b}{a_P} \right)^2.
\end{eqnarray} 
Here, we have neglected the Coulomb logarithm $\ln \Lambda$, which is of order $\ln \Lambda \sim 10$.  Using this factor, and combining Eqs. (\ref{eq:delta_a}--\ref{eq:Nt}), we find that
\begin{eqnarray}
	\frac{\langle (\delta a)^2 \rangle}{a^2} \sim 10 \left( \frac{M_P}{M_\odot} \right)^2 \left( \frac{a}{a_P} \right)^2 \frac{t}{P_\chi}.
\end{eqnarray}

Let us consider the time for $\langle (\delta a) \rangle / a^2 = 1$ for each of the inner planets.  For Mercury, $a_{\text{\mercury}} \approx 0.4 \hbox{ AU}$ and $M_{\text{\mercury}} / M_\odot \approx 10^{-7}$.  Thus, for a WIMP with $a  = 1\hbox{ AU}$, it takes $\sim 10^{12}$ yr for the WIMP semi-major axis to change significantly.  For Venus and Earth, $M_P/M_\odot \sim 10^{-6}$ and $a \sim 1\hbox{ AU}$.  The timescale for significant changes to a WIMP orbit with an initial semi-major axis of $a = 1\hbox{ AU}$ is of order $10^{10}$ yr.  Mars has a smaller mass ($M_{\text{\mars}} \approx 0.1 M_\oplus$) and larger semi-major axis ($a_{\text{\mars}}\approx 1.5\hbox{ AU}$) than the Earth, so the timescale for WIMP orbits to change due to encounters with Mars is much longer than from interactions with the Earth or Venus.  All of these timescales are longer than the age of the solar system, the maximum time for which we consider WIMP orbits, so it is unlikely that a large number of WIMPs with $a < 2.6 \hbox{ AU}$ have thermalization properties that are different from those discussed in Section \ref{sec:insight}.

Now that we have mapped out the annihilation rate suppression in WIMP parameter space, it is possible to determine how serious the suppression may be for current and future neutrino telescopes.  For this discussion, we only consider changes to limits on \sigmapsd$\,$ since (i) we showed in Section \ref{sec:insight} that there is comparatively little suppression to the annihilation rate if spin-independent interactions dominate in the Sun and (ii) given existing limits on \sigmapsi, current and planned neutrino telescopes cannot detect WIMP annihilation in the Sun driven by spin-independent scattering \cite[e.g.,][]{angle2008,cdms2008}.

First, we consider the existing limits on \sigmapsd$\,$ from the Super-Kamiokande experiment \cite{desai2004}.  In the analysis of their data, neutrino oscillations were neglected but limits on the cross section were otherwise derived using fairly conservative assumptions for a supersymmetric WIMP.  The most stringent limit on the cross section came for $m_\chi \sim 100\hbox{ GeV}$, \sigmapsd$\: \lesssim 10^{-39} \hbox{ cm}^2$ \cite{desai2004}.  For larger masses, the limit on \sigmapsd$\,$ scales as $m_\chi$ because (i) for such cross sections, the annihilation rate will have reached equilibrium, and so $\Gamma_a \propto C \propto \sigma_p^{SD} m^{-2}_\chi$; and (ii) the number of neutrinos produced in a single WIMP annihilation event scales approximately as $m_\chi$ \cite{kamionkowski1991}.  From Fig. \ref{fig:param} in the previous section, it appears unlikely that the limit on \sigmapsd$\,$ from Super-Kamiokande is greatly suppressed.  If \sigmapsd$\:$ were right at the flux limit for $m_\chi = 100 \hbox{ GeV}$, most WIMPs on Jupiter-crossing orbits would be ejected before rescattering;  however, only $f_\text{\jupiter} \sim 10^{-2}$ are Jupiter-crossing.  From Fig. \ref{fig:param}, we see that for WIMPs with higher masses, even Jupiter-crossing WIMPs should thermalize, so the annihilation rate of WIMPs in the Sun should be well-described by Section \ref{sec:std}.

Next-generation neutrino telescopes are anticipated to have $\sim 100$ times the sensitivity of Super-Kamiokande to neutrinos from WIMP annihilation in the Sun \cite{delosheros2008}.  If we were to naively scale the limits on \sigmapsd$\:$ from Super-Kamiokande to km$^3$-scale experiments, we would find \sigmapsd$\: \lesssim 10^{-41} \hbox{ cm}^2$ for $m_\chi = 100\hbox{ GeV}$ and \sigmapsd$\: \lesssim 10^{-39}\hbox{ cm}^2$ for $m_\chi = 10\hbox{ TeV}$.  In reality, from Section \ref{subsec:map}, we find that the limits on \sigmapsd$\:$ should be weaker by a factor of $(1-f_\text{\jupiter})^{-1}$ (Fig. \ref{fig:frac}).  Alternatively, this means that the next-generation neutrino experiments will be sensitive to a smaller range of \sigmapsd$\:$ than currently predicted by the experimental collaborations.  The restriction is only significant (a change in the limit by $> 10\%$) if $m_\chi \gtrsim 1\hbox{ TeV}$.

It should be noted that the interpretation of an observed neutrino flux, or any attempt to map the observed flux to an elastic scattering cross section, will need to include the effects of the annihilation branching fractions and neutrino oscillation \cite{blennow2008,cirelli2005,barger2007b,lehnert2008}.  These latter effects may complicate the estimate of the annihilation rate in the Sun.

One may also consider the prospects of observing WIMP annihilation in the Sun for specific WIMP models:

\emph{Supersymmetry}:  Even in limited scans of the MSSM, it is apparent that a large range of \sigmapsd is allowed, up to \sigmapsd$\sim 10^{-38} \hbox{ cm}^2$ \cite{gondolo2004}.  Therefore, it is possible that MSSM neutralino annihilation may be observed with next-generation neutrino telescopes (e.g., Antares, IceCube).

\emph{Universal Extra Dimensions}:  In the minimal version of this model (only one extra dimension), \sigmapsd$\: \gg \:$\sigmapsi, so we consider only the prospects for observing neutrinos if spin-dependent interactions dominate the capture rate of WIMPs in the Sun.  The expected spin-dependent cross section goes as \cite{hooper2007}
\begin{eqnarray}
	\sigma_p^{SD} \approx 1.8\times 10^{-42} \left( \frac{1\hbox{ TeV}}{m_\chi}\right)^4 \left(\frac{0.1}{\Delta}\right)^2 \hbox{ cm}^2,
\end{eqnarray}
where $\Delta$ is the fractional mass difference between the WIMP (the Kaluza-Klein photon) and the Kaluza-Klein quark.  In order to satisfy relic abundance requirements, $m_\chi \gtrsim 500\hbox{ GeV}$, although the exact lower limit depends on boundary terms in the UED Lagrangian \cite{hooper2007}.  The allowed range of $m_\chi - $\sigmapsd parameter space straddles the line between cases (a2) and (a3) from Section \ref{subsec:map}, and also straddles the critical number equilibrium timescale, $t_e \sim t_\odot$.  Both the high WIMP escape fraction and the possibility that the number of WIMPs in the Sun is not in equilibrium drive the annihilation rate down if $m_\chi \gtrsim 1 \hbox{ TeV}$. It would take a telescope with at least an order of magnitude more sensitivity than IceCube to detect even the Kaluza-Klein photon with the best prospects for detection.

\emph{Little Higgs}:  As in UED models, \sigmapsd$\: \gg \:$\sigmapsi, so again we consider only the prospects of finding neutrinos from WIMPs captured in the Sun by spin-dependent interactions \cite{birkedal2006}.  In Little Higgs models with $T$-parity, the natural scale for the heavy photon, the WIMP in this model, is $m_\chi < 500\hbox{ GeV}$, and the spin-dependent cross section scales with mass as
\begin{eqnarray}
	\sigma_p^{SD} \sim 5\times 10^{-47} \left( \frac{ 1 \hbox{ TeV}}{m_\chi} \right)^4 \left( \frac{0.1}{\Delta} \right)^2 \hbox{ cm}^2,
\end{eqnarray}
where $\Delta$ is the fractional mass difference between the heavy photon and the $T$-odd quark \cite{cheng2003,birkedal2004,birkedal2006,hubisz2005,hooper2007c,perelstein2007}.  For a fiducial case of $\Delta = 0.1$, \sigmapsd$\: \sim 5\times 10^{-43}\hbox{ cm}^2$ for $m_\chi = 100\hbox{ GeV}$ and \sigmapsd$\: \sim 8 \times 10^{-46}\hbox{ cm}^2$ for $m_\chi = 500\hbox{ GeV}$.  In the former case, the number of WIMPs in the Sun will marginally be in equilibrium, and the suppression due to WIMP populations with $a > 1.5 \hbox{ AU}$ will be negligible.  One would still require $\sim 10$ times the sensitivity of IceCube to detect such a WIMP.  In the latter case, the number of WIMPs in the Sun is small and far from equilibrium, but suffers little suppression due to WIMPs with $a > 1.5\hbox{ AU}$.  However, if $\Delta$ is not much larger than the fiducial value, $X P_\chi / \tau \gtrsim t_\odot$ for the initial median semi-major axis of captured WIMPs, and so the WIMP annihilation rate in the Sun will quickly drop to almost nothing.  Therefore, one would expect virtually no neutrino signal from the Sun for a reasonable swath of Little Higgs parameter space.

\begin{acknowledgments}
This work grew out of a Ph.D thesis completed at Princeton University.  We would like to thank Scott Tremaine for patient advising, keen insight, and comments on the draft.  We thank A. Serenelli for providing me with the standard solar model in tabular form.  We acknowledge financial support from NASA grants NNG04GL47G and NNX08AH24G and from the Gordon and Betty Moore Foundation. The simulations were performed using computing resources at Princeton University supported by the Department of Astrophysical Sciences (NSF AST-0216105), the Department of Physics, and the TIGRESS High Performance Computing Center.
\end{acknowledgments}


\end{document}